\documentclass[aps,prl,twocolumn,nofootinbib,superscriptaddress]{revtex4-2}

\usepackage{graphicx}
\usepackage{hyperref}
\usepackage{physics}
\usepackage{bm}
\usepackage[utf8]{inputenc}
\usepackage{amsmath,amssymb,amsthm}
\usepackage{enumitem}
\usepackage{natbib}

\begin{document}

\title{A Lorentzian FRG Investigation of the Quasi--Static Weak--Field Infrared Limit of Gravity}

\author{Krzysztof Nowak}

\email{knowak@wz.uw.edu.pl}
\affiliation{University of Warsaw, Faculty of Management}

\date{\today}

\begin{abstract}

A common assumption in the Effective Field Theories of gravity is that their quasi--static weak--field infrared limit yields the well--known second--order Poisson operator. We examine this limit for the universality class of parity--even, symmetric, analytic gravitational theories admitting a local derivative expansion using Lorentzian FRG methods. We find that, in the curvature--squared truncation, the scalar--trace sector self--closes at $\mathcal{O}(q^4),$ allowing the projected flow to be obtained by analytic continuation of the corresponding Euclidean result. This yields a screened d'Alambertian $D_\ell \equiv (1+\ell^2 \Box)\Box$ characterised by an emergent correlation length $\ell$. We show the operator is consistent with the ADM constraint structure and thus it does not introduce propagating scalar ghosts in the scalar--trace sector. We further derive its retarded response kernel and show its static--limit Green's function in response to a point source, which reduces to Newtonian gravity for $\ell \rightarrow 0$.

\end{abstract}

\maketitle

\section{Introduction}

The infrared (IR) limit of Effective Field Theories (EFT) treatments of gravity is commonly assumed to be trivial: in the quasi--static weak--field regime, the gravitational response reduces to the second--order Poisson operator yielding the familiar $1/r$ Newtonian potential. Deviations from this form are typically attributed to additional fields or modifications of gravity. Despite, this expectation has not been derived from first principles using the Functional Renormalisation Group (FRG) framework in Lorentzian signature. Here we challenge this expectation by deriving the quasi--static weak--field IR endpoint in a broad local metric universality class, finding a screened d’Alembertian structure in the scalar--trace sector.

FRG methods in this regime face two technical obstacles. Firstly, even if the EFT is analytic, it generally contains an infinite number of derivative expansion terms with mixing between scalar, vector, and tensor terms, so one needs a projection that closes at a given order. Secondly, such an analysis needs to be performed in Lorentzian, not in the more familiar Euclidean signature ~\citep{reuter1998nonperturbative, fehre2023lorentzian}. This necessarily introduces issues related to causal structure, retardation, and Ostrogradsky instabilities.

Our analysis is motivated by derivative expansions and curvature--invariant truncations commonly employed in FRG approaches in asymptotic safety \citep{Codello:2008vh,Falls:2016msz,Denz:2018cql}. In Section II we show that in a broad universality class of theories of gravity the scalar sector projection self--closes at $\mathcal{O}(q^4),$ where $q$ is the norm of the momentum vector, in the quasi--static weak--field IR limit. This allows a solution of the Wetterich equation in this truncation through analytic continuation via Wick rotation ~\citep{manrique2011asymptotically} of the Euclidean signature solution (using the equations derived in \citep{Benedetti2010}) as shown in Section III. This yields the paper's main result:  IR endpoint corresponds to a hyperbolic operator, a screened d'Alambertian $D_f \equiv (1+\ell \Box) \Box$ whose behaviour is controlled by an emergent correlation length $\ell$. 

In section IV, using the example of ADM decomposition we show that our construction does not introduce Ostrogradsky ghosts in the scalar--trace sector of the truncation, as it preserves the constraint structure of the underlying gravitational theory ~\citep{ADM1962,HenneauxTeitelboim1992}. Finally, we derive the retarded response kernel. In the static limit, this yields an exponentially screened Poisson Green's function, which for $\ell \rightarrow 0$ reproduces Newtonian gravity.

\section{Setup and Assumptions}
We work in the standard Functional Renormalization Group (FRG) framework using the effective average action $\Gamma_k$ at a coarse-graining scale $k$, focusing on the quasi-static weak-field IR limit. Throughout, we employ the Lorentzian signature $(-+++)$ with $\Box = - \partial^2 _t + \nabla^2$, and in momentum space  $p_\mu=(\omega,\mathbf{q})$ such that

\begin{equation}
\Box \rightarrow p^2 \equiv -\omega^2 +q^2,
\end{equation}
where $q^2$ denotes the positive eigenvalue of $-\nabla^2$ and analytic continuation is performed via the Wick rotation $p^2_E \rightarrow -p^2$. The static projection is $\Box \rightarrow \nabla^2$.  

We consider a universality class of gravitational theories defined by the following assumptions:

\begin{enumerate}
    \item \textbf{Parity--even, symmetric, torsion--free metric theory.}  
    The effective average action involves only parity--even curvature invariants built from the metric, with no torsion or Chern--Simons-like terms.

    \item \textbf{Local analytic derivative expansion.}  
    The effective action $\Gamma_k[g]$ admits a local analytic expansion in curvature invariants and their covariant derivatives:
    $ \Gamma_k[g] = \sum_{n \ge 0} c_{n,k}\,\nabla^{2n} R + \cdots,$
    with no non-analytic (e.g. fractional or nonlocal) structures at the scales of interest (as in ~\citep{Codello:2008vh}).

    \item \textbf{Weak-field background split.}  
    We work in the linear background-field expansion
    $g_{\mu\nu} =  \eta_{\mu\nu} + h_{\mu\nu},$
    appropriate for the quasi-static weak--field regime, and assume sources couple to the potential $\Phi$ linearly.

    \item \textbf{Quasi-static scaling.}  
    Temporal variations are parametrically small compared to spatial gradients:
    $  |\omega|/q \equiv \epsilon \ll 1.$ Equivalently $\partial_t \sim \epsilon \nabla,$ so $\partial_t 
    \sim \epsilon q$ while spatial gradients scale as $q$,
    ensuring that spatial derivatives dominate the IR flow of the scalar trace. This limit is commonly used in gravitational perturbation theory and effective field approaches to modified gravity~\citep{Koyama:2015vza}
    
    \item \textbf{Truncation stability}
    We assume that after projection onto the scalar trace sector in the quasi--static weak--field limit the Lorentzian FRG flow preserves the derivative hierarchy of the truncation up to $\mathcal{O}(q^4)$, so that operators of higher derivative order do not contribute to the projected flow of the scalar couplings retained at this order.
    
\end{enumerate}

Given these assumptions, any curvature invariant of order $n$ contributes only at order $q^{2n}$ in the scalar trace or reduces to the $R^2$ and $R_{\mu \nu}R^{\mu \nu}$ structures modulo topological terms. Correspondingly, for the purpose of our argument, it is sufficient to consider the curvature squared truncation of the gravitational action:
\begin{equation}
\Gamma_k^{\rm} = \int d^4x \sqrt{-g}
\bigl[\frac{1}{2} Z_k (R - 2\Lambda_k) + a_k R^2 + b_k R_{\mu\nu}R^{\mu\nu}\bigr].
\end{equation}

We use the background--field method with de Donder--type gauge fixing and Faddeev--Popov ghosts \citep{Buchbinder:1992rb,Reuter:1996cp}. In the quasi--static IR, the contributions of these sectors can be absorbed into the definitions of the constants and do not contribute to the scalar kernel in any other way at order $\mathcal{O}(q^4)$; they will therefore be omitted. We also omit the analysis of vacuum energy and set $\Lambda_k = 0$ for this derivation, leaving its treatment to future work.
Throughout, the notation $k \rightarrow 0$ denotes the infrared endpoint of the FRG flow, where the regulator is removed, and the effective two--point operator freezes. 

\section{IR closure of the scalar sector}

To assess the IR response in the quasi--static regime, we employ the quasi--static scaling and perform an expansion in spatial gradients $|{\bm q}| = q$ at fixed $\epsilon$.
Projecting the quadratic contributions of the $R\), \(R^2\) and \(R_{\mu\nu} R^{\mu\nu}$ onto the scalar trace  $h = h^\mu_{\ \mu}$ mode by mode yields: 

 \begin{equation}
\Gamma_{\text{k, scalar}}^{(2)} (q) 
\sim \frac{Z_k}{4} q^2 +  \frac{\alpha_k}{2} q^4  + \mathcal{O}(q^4 \epsilon, q^6).
\label{eq:scalar-scaling}
\end{equation}
Here $Z_k$ is the running wave--function renormalisation and $\alpha_k \equiv 3a_k + b_k$ is the scalar coupling. Vector and tensor contributions are of order  $\mathcal{O}(q^4 \epsilon)$ or $\mathcal{O}(q^6)$ respectively, and consequently do not back--react onto the scalar trace sector at the order displayed. Consequently, the scalar trace sector self--closes at order $\mathcal{O}(q^4)$. The equations used to extract the scalar trace are presented in Appendix ~A and the explicit derivative counting is presented in Appendix ~B.

This hierarchy arises from the dominant contributions provided by the quadratic curvature invariants in the quasi--static limit. The Einstein--Hilbert term contributes two spatial derivatives through ($h\,\nabla^{2}h$) to the scalar--trace and the curvature--squared invariants contribute four derivatives of the form $h\nabla^{4}h$. The mixed scalar--vector term contains at least one extra derivative suppressed by $\epsilon$ acting on the divergence--free vector coming from the requirement of a time derivative $\partial_t$ required for a non--vanishing back--reaction on the scalar sector. The transverse--traceless tensor enters only at $\mathcal{O}(q^{6})$ as two derivatives are required for the $TT$ projection, two are required for projection onto the scalar trace, and an additional two are required to enter the kinetic operator. Mixed scalar--vector or scalar--tensor terms contain at least one additional derivative acting on the divergence--free vector or transverse--traceless tensor, because these modes enter only through projected derivatives. Consequently, the vector sector first enters at scale $\mathcal{O}(q^{4}\epsilon)$ and the tensor sector at $\mathcal{O}(q^{6})$. Thus, although these modes are present, they are suppressed in the quasi--static IR limit.

Beyond the local curvature--squared truncation, the effective average action will contain higher--order derivative terms such as $R_{\mu\nu\rho\sigma}R^{\mu\nu\rho\sigma}$, $R\Box R$, and similar structures, as well as nonlocal structures generated by integrating out fluctuations at intermediate scales. Such terms contribute to the scalar sector at orders $\mathcal{O}(q^6)$ for operators with six or more derivatives, or order $\mathcal{O}(q^4)$, but in combinations that can be re--expressed in terms of $R^2$ and $R_{\mu\nu} R^{\mu\nu},$ plus terms that are suppressed in the scalar--trace sector. This follows from the fact that, in four dimensions, the Gauss Bonnet combination $E \equiv R_{\mu\nu\rho\sigma}R^{\mu\nu\rho\sigma}-4R_{\mu\nu}R^{\mu\nu}+R^2$ is a topological invariant at the classical level and does not contribute to the local equations of motion in the weak--field expansion ~\citep{Lovelock:1971yv}. Consequently, any $R_{\mu\nu\rho\sigma}$ term can be re--expressed as a combination of $R_{\mu\nu} R^{\mu\nu}$ and $R^2$ plus a topological term that does not contribute to the local equations of motion in the weak--field expansion ~\citep{Lanczos1938}. Similarly, total--derivative terms like $\Box R$ can be removed by integrations by parts or redefinitions of the scalar coupling in the quadratic kernel, up to boundary contributions that are irrelevant for the bulk Green’s functions. 

Consequently, the scalar trace sector self--closes at $\mathcal{O}(q^{4})$ and thus for the curvature--squared truncation within our universality class in the IR limit we focus on the running couplings  $Z_k$ and $\alpha_k$ in this projection. We do not claim that the quasi--static hierarchy is dynamically generated or preserved by the full RG flow.

\section{Lorentzian Endpoint and Effective Operator}

The scale dependence of the effective action is governed by the Lorentzian Wetterich equation \cite{Wetterich1993}. Our result is obtained by starting the Euclidean Wetterich equation
\begin{equation}
\partial_k \Gamma_k 
= \frac12 \mathrm{Tr}\!\left[ \left(\Gamma_k^{(2)} + R_k\right)^{-1} 
\partial_k R_k \right] ,
\label{eq:wetterich}
\end{equation}

and performing a Wick rotation to obtain the Lorentzian formulation. We use the analytic Type-II (spectrally adjusted) regulator
\begin{equation}
R_k(p) = \frac{Z_k}{8} k^2 r(p^2 / k^2),
\label{eq:regulator}
\end{equation}
where $r$ is an analytic shape function satisfying the usual FRG conditions.  

The flow of $Z_k$ and $\alpha_k$ is analysed by considering the Euclidean FRG of the corresponding truncated effective average action. To control the IR behaviour of the couplings $(Z_k,a_k,b_k)$ we evaluate the Euclidean functional renormalisation group flow in the curvature--squared truncation using the background field method ~\citep{Benedetti2010}. The Euclidean flow is evaluated on a generic Einstein background
$\bar R_{\mu\nu}=\tfrac{1}{4}\bar R\,\bar g_{\mu\nu}$ using a Type II cutoff. While we do not fix the shape function explicitly, the existence and structure of the IR endpoint, and the resulting scalar operator, are robust across standard choices. 

On an Einstein background, the quadratic curvature invariants are not independent.
Using the Gauss--Bonnet identity, its Weyl--decomposed form using the trace-free Weyl tensor $C_{\mu\nu\rho\sigma},$ their relationship is:
\begin{equation}
R_{\mu\nu}R^{\mu\nu}
= \frac12 C_{\mu\nu\rho\sigma}C^{\mu\nu\rho\sigma}
+ \frac13 R^2 - \frac12 E.
\end{equation}

Accordingly, the curvature--squared truncation can be rewritten, up to the topological
term $E$, in terms of two independent couplings
\begin{equation}
\bar a_k \equiv a_k + \frac{b_k}{3},
\qquad
\bar b_k \equiv \frac{b_k}{2}.
\end{equation}
The scalar trace sector depends only on the combination
\begin{equation}
\alpha_k \equiv 3a_k + b_k = 3\bar a_k ,
\label{eq:alpha-def}
\end{equation}
so that $\beta_{\alpha}=3\,\beta_{\bar a}$ in the background approximation.

Introducing dimensionless couplings
\begin{equation}
g_1 \equiv -\frac{Z_k}{  k^2}, \qquad
g_2 \equiv \bar a_k, \qquad
g_3 \equiv \bar b_k ,
\end{equation}

We adopt the curvature--squared Euclidean background-flow system as derived in ~\citep{Benedetti2010} for the Type-II cutoff: 
\begin{align}
\partial_t g_1 &= -2 g_1 + \frac{1}{2(4\pi)^2}
\Big[ C_2 + \tilde C_{e2}
-\frac{7}{3}\Phi^1_1(0) - \frac{2}{3}\Phi^2_2(0) \Big], \label{eq:beta-g1}\\
\partial_t g_3 &= \frac{1}{(4\pi)^2}
\Big[ \frac{1}{360}C_3 + \frac{5}{9}\tilde C_{e3}
+ \frac{11}{360}\varphi \Big], \label{eq:beta-g3}\\
\partial_t g_2 - \frac16 \partial_t g_3 &=
\frac{1}{(32 \pi)^2}
\Big[ \frac12 C_4 + \frac12 \tilde C_{e4}
-\frac{1}{18}\Phi^1_2(0) - \frac{1}{9}\Phi^2_3(0)
-\frac{1}{15}\varphi \Big], \label{eq:beta-g2}
\end{align}
where $C_i$ and $\tilde C_{ei}$ denote the scalar and transverse--traceless
contributions to the flow, expressed in terms of generalised threshold functions,
and the quantity $\varphi$ collects universal curvature--squared terms.
The system \eqref{eq:beta-g1}--\eqref{eq:beta-g2} is linear in the beta functions and
can be solved algebraically at each point in coupling space for standard regulator choices. Although these flows are used to analyse UV fixed points in the original analysis, the projected beta functions define a scale evolution that can equally be integrated toward the IR endpoint within the same truncation; the limitation is that the truncation’s quantitative reliability in the deep IR is not guaranteed. 

Using the definitions above, the running of the original couplings follows from
\begin{equation}
\beta_Z = - k^2\big(\beta_{g_1}+2g_1\big),
\qquad
\beta_{\alpha}=3\,\beta_{g_2}.
\end{equation}
After projection, the flat--background limit $\bar R\to 0$ is taken, yielding the
Euclidean flow relevant for the Lorentzian quasi--static sector. Though, the explicit running of the couplings are regulator--dependent, the functional form is not. This is because the scalar kernel is fixed by locality and at $\mathcal{O}(q^4)$ there are only two available operators in this sector: $p^2$ and $p^4$. Thus, the regulator dependence is expected to enter only through the numerical value of the emergent correlation scale:

\begin{equation}
\ell^2 \equiv \frac{2 \alpha_{k}}{Z_{k}} \Big|_{k \rightarrow 0}.
\end{equation}
Where $\ell$ is in units of length.

We set the normalisation by taking the limit $Z_{k\to 0} \equiv  \frac{1}{8 \pi G}.$ The resulting IR endpoint scalar kernel in Euclidean signature, truncated to the terms that remain relevant in the IR limit, takes the form:
\begin{equation}
f_k(p_E^2) = \frac{1}{32 \pi G}(p_E^2 + \ell^2 p_E^4),
\label{eq:fk-euclidean}
\end{equation}
where the squared momentum in Euclidean signature is $p_E^2 = \omega_E^2 + q^2$. 

We use a Wick rotation method to obtain the corresponding Lorentzian endpoint solution ~\citet{manrique2011asymptotically}. While generic Lorentzian FRG flows generate branch cuts and logarithms associated with multi--particle continua (see ~\citep{}), in our case, they are projected out as our truncation is local and contains only polynomials in $p_E^2$. Thus, the analytic continuation of the quadratic kernel to Lorentzian signature is unique via the Wick rotation within this truncation. 

Writing 
\begin{align}
p_E = (\omega_E,\bold{q}), p_L = (\omega,\bold{q}), \omega_E=i \omega,
\end{align}
the Wick rotation corresponds to

\begin{align}
p_E^2 \mapsto -p_L^2 = \omega^2 - q^2.
\end{align}

Because at the endpoint $f_{k \rightarrow 0}(p^2_E)$ is an analytic polynomial in $p_E^2,$ the Wick rotation is consistent with the quasi--static truncation. In Euclidean signature, the quasi--static hierarchy enforces $\omega^2_E \ll q^2$, which is preserved under the Lorentzian continuation. We therefore perform the Wick rotation after projection, without introducing additional ambiguities in the truncated scalar sector. The resulting Lorentzian endpoint kernel is:

\begin{equation}
\Gamma^{(2)}_{k \rightarrow 0}(p) = \frac{1}{32 \pi G} p^2(1 + \ell^2 p^2).
\label{eq:IR-operator-momentum}
\end{equation}

Equivalently, in position space, up to a scaling constant, it takes the form of a Lorentzian operator which we refer to in the remaining text as the screened d'Alambertian:

\begin{equation}
D_\ell \equiv (1 + \ell^2 \Box)\Box,
\label{eq:IR-operator}
\end{equation}

Correspondingly, the  scalar effective action can be written as
\begin{equation}
S_{k \rightarrow 0}[\Phi]
= \frac{1}{2} \int d^4x\sqrt{-g}\;
\Phi \frac{1}{32 \pi G} D_\ell \Phi -
\frac{1}{8}\int d^4x \sqrt{-g} \Phi \rho .
\label{eq:effecive_action}
\end{equation}

The regulator in \eqref{eq:regulator} is Wick--rotated together with the kernel, and the retarded boundary conditions enter only through the rotated prescription of the response kernel. The full Lorentzian two--point function in the scalar sector is obtained by replacing $p^2_E \rightarrow -p^2_L$ in the Euclidean kernel so that: 
\begin{equation}
\Gamma^{(2)}_k(p) + R_k(p) \equiv f_k(-p^2_L)+R_k(p).
\end{equation}

At the endpoint, it yields the Lorentzian retarded response kernel:
\begin{equation}
G_{\mathrm{ret}}(p)
=  32 \pi G \left[\frac{1}{p^{2}-(\omega+i\varepsilon)^2}
       - \frac{1}{p^{2}+\ell^{-2}-(\omega+i\varepsilon)^2}\right]
\label{eq:Gretagain}
\end{equation} 

where $\varepsilon$ denotes the standard $i \varepsilon$ perception. The pole structure is precisely the Lorentzian continuation of the endpoint Euclidean kernel  ~\eqref{eq:fk-euclidean}. Since both the kernel and the regulator are analytic functions of $p^2$, the Wick rotation does not cross any singularities; consequently the continuation preserves the causal pole structure and keeps the poles below the real axis in the retarded prescription. Thus, the Lorentzian endpoint  is characterised by the same values ($Z_{k \rightarrow 0}, \alpha_{k \rightarrow 0}$) and thus the same emergent correlation scale $\ell$. 

\section{Propagating Scalar Ghosts and Causality}

The screened d'Alambertian $D_\ell$ has a corresponding retarded response kernel given by ~\eqref{eq:Gretagain} which contains a pair of poles with opposite residues. In Stelle gravity, which is an explicit curvature--squared theory of gravity, this leads to the Ostrogradsky instabilities ~\citep{Stelle1977}. One of the few ways to deal with this issue, in higher--derivative gravitational theories, is by examining the constraint structure ~\cite{Chen2013}. Constraints work by exploiting the fact that Ostrogradsky's theorem requires non-degenerate higher--time--derivative Lagrangians for true dynamical variables as outlined in ~\citep{woodard2007avoiding}. This is exactly our case as the scalar--trace mode remains a constrained, non-dynamical variable. 

To illustrate this mechanism, we consider ADM decomposition and its constraint structure~\citep{ADM1962}. The ADM split we decompose the space--time metric $g_{\mu \nu}$ using lapse $N$ and shift $N^i$ where $i$ and $j$ are spatial indices the following way:

\begin{equation}
d s^2 = - d t^2 + g_{i j} (dx^i +N^i dt)(dx^j + N^j dt).
\end{equation}

This allows for examining space slices with coordinates $x^i$ as slices indexed by a time coordinate $t$, where slices are parametrised via $N$ and $N^i$. In the ADM formalism the lapse $N$ and shift $N^i$ act as Lagrange multipliers enforcing the Hamiltonian and momentum constraints. 

For the FRG flow to acquire canonical momenta in our construction, and thus enlarge the canonical phase space of the original theory, there needs to be a time derivative acting on the lapse and shift \citep{dirac1950generalized}. However, in our case in the quasi--static weak--field regime, coarse--graining at scale $k$ we preserve these constraints at $\mathcal{O}(q^4),$ in the scalar--trace projection, as time derivatives enter first at $\mathcal{O}(\epsilon q^4)$ sourced by the scalar--vector mixing term (see Appendix ~B). Thus, although the operator $D_\ell$ is hyperbolic, its action on the scalar trace does not convert the constraint equation into a genuine evolution equation. This situation differs from higher--derivative theories such as Stelle gravity, where additional time derivatives appear prior to constraint reduction and enlarge the canonical phase space ~\citep{Stelle1977}. 

Thus, although in the quasi--static IR limit, the effective response equation sourced by $\rho$, retaining terms up to $\mathcal{O}(q^4)$ is:

\begin{equation}
-\partial_t^2\Phi+\nabla^2\Phi+\ell^2\nabla^4\Phi  = 4 \pi G \rho
\label{eq:Motion}
\end{equation}

this equation should be interpreted as an equation determining $\Phi$ subject to causal retardation, but not defining an independent Cauchy problem requiring initial data. Consequently, the second pole contributes only to a causal, retarded response in the scalar Green’s function. It produces exponential screening at the scale $\ell,$ and does not correspond to a physical particle or propagating mode. A full non--perturbative Hamiltonian analysis of the Lorentzian FRG effective action lies beyond the scope of this work, but is not required for the sectoral claim made here.

\section{The static limit Green's functions}

The response $G(r)$ to point sources in the static limit corresponding to $\partial_t \rightarrow 0$, and requiring it to obey $G(r \rightarrow \infty) \rightarrow 0$ in asymptotically flat space, we obtain the Green's function:
\begin{align}
G(r) &= \frac{1-e^{-r/\ell}}{4\pi r}.
\end{align}

We further notice it reduces to the standard Poisson Green's function in the limit $\ell \rightarrow 0$.

\section{Discussion and Conclusions}

We derived the quasi-static weak-field IR endpoint operator, which we refer to as the screened d'Alambertian $D_\ell$, for a large universality class of gravity theories using Lorentzian FRG. We have shown scalar--sector closure at $\mathcal{O}(q^4)$. The operator is characterised by a single emergent correlation length $\ell$ and this single scale governs the gravitational potential in the scalar sector in the relevant limit. In contrast to the typically assumed Newtonian $1/r$ behaviour, the resulting static potential exhibits exponential screening set by $\ell,$ implying a finite gravitational potential energy.

Our results indicate there is a single correlation length $\ell$ in the deep IR that governs the gravitational EFT set by the asymptotic behaviour of the ratio $\ell^2=\alpha_k/Z_k \Big|_{k \rightarrow 0}$. While we expect the functional for of the IR fixed--point operator to be robust in our universality class, its numerical value may be context--dependent. For example, matter fields are known to generally modify the beta functions of the running couplings $Z_k$, $\alpha_k$ \cite{Benedetti2010}.

The screened d’Alembertian is not exotic in form. Similar, higher--derivative operators, have appeared previously in curvature--expanded theories of gravity, including Stelle--type theories of gravity and $f(R)$ frameworks developed in the FRG approach  ~\citep{Stelle1977, Codello:2008vh}. They generally, however, require external constraints or new fields to deal with Ostrogradsky instabilities. The novelty of the present result lies the emergence of the operator as a universal IR endpoint result, derived from first principles within Lorentzian FRG, without introducing additional degrees of freedom or modifying the underlying gravitational theory. We note, that the exponential screening in the static limit Green's function arises in a different mechanism than in Yukawa gravity theories where it follows from a second--order condition and yields Green's functions with Yukawa exponential screening of a different functional form ~\citep{Will:2014kxa}.

In our analysis, the screened d’Alembertian does not introduce additional propagating degrees of freedom. Its structure is consistent with the ADM constraint analysis of General Relativity, in that the lapse and shift remain non--dynamical, as long as the underlying theory is already well behaved, and the Hamiltonian and momentum constraints are preserved. Thus, the scalar trace continues to act as a constrained variable rather than an independent mode and is compatible with causal Lorentzian evolution in the quasi--static weak--field regime.

We note some of the limitations of our results. Firstly, our results apply only in the universality class considered. Explicitly, our results do not apply to theories in which non--locality forbids a local analytic derivative expansion, to theories with strong underlying anisotropy, or to non–parity-even or non-symmetric theories whose symmetry structure modifies the Wetterich-derived beta functions. In our analysis, we also explicitly assumed linear coupling of density to potential in the weak field regime. We also assumed expansion around flat (Minkowski) background, and we leave the derivation of the equations in the FLRW background for future work. Lastly, in our analysis, we explicitly left out vacuum sources, setting $\Lambda_k = 0$. If vacuum source runnings are shown to be strongly dependent on the coarse-graining scale $k$ the validity of our assumption affects our results. We leave the treatment of vacuum sources under Lorentzian FRG to future work. 

Future work is needed to extend the present analysis beyond a flat Minkowski background. In particular, deriving the quasi-static IR endpoint structure on cosmological backgrounds such as FLRW spacetimes and assessing the role of vacuum sources with $\Lambda_k \neq 0$ would clarify the robustness of the screened d'Alambertian in more general settings. In addition, a more complete understanding of the emergent correlation scale $\ell$ would be obtained by incorporating matter and dark--matter field contributions dynamically into the Lorentzian FRG. 

More broadly, our analysis challenges the long--held assumption that the EFT of General Relativity itself remains Newtonian in the quasi-static weak-field IR limit. Using Lorentzian FRG analysis, we have shown that within parity--even, local gravitational theories in our universality class, this is not the case. The IR fixed--point scalar sector exhibits a screened d'Alambertian structure, leading to exponential screening in the static Green's function. This behaviour does not rely on modifying the degrees of freedom of General Relativity, but arises from the structure of the effective action under coarse--graining. As a result, the screened IR response is compatible with a wide class of gravity frameworks whose effective actions fall within the assumed universality class.

\bibliographystyle{plainnat}
\bibliography{final_paper_draft_ref}

\appendix

\section*{Appendix A: Quadratic weak--field expansion and scalar projection}
\label{app:scalar_projection}

\setcounter{equation}{0}
\renewcommand{\theequation}{A.\arabic{equation}}

In this Appendix we present the minimal weak--field expansions required to
extract the scalar (trace) part of the quadratic kernel in the quasi--static
regime. We determine the form of the scalar operator and
identify the coupling combination controlling the leading four--derivative contribution in the scalar--trace projection up to gauge--dependent terms that do not contribute to the scalar trace.

\subsection{Background split and conventions}

We work with the weak--field expansion $g_{\mu\nu} = \eta_{\mu\nu} + h_{\mu\nu}$, where the background is Minkowski with conventions as in the main text. Accordingly, indices are raised and lowered with $\eta_{\mu\nu}$, and covariant derivatives reduce to partial derivatives. We denote the scalar trace by $h$. 

The Ricci scalar is expanded as
\begin{equation}
R = R^{(1)}[h] + R^{(2)}[h] + \mathcal{O}(h^3),
\end{equation}
with
\begin{equation}
R^{(1)} = \partial_\mu \partial_\nu h^{\mu\nu} - \Box h .
\end{equation}
Only the pieces contributing to the scalar trace sector after projection, discarding total derivatives, and gauge–dependent terms will be retained below. Since $\sqrt{-g} = 1+\frac{1}{2} h+\mathcal{O}(h^2)$, at quadratic order we set $ \sqrt{-g} \rightarrow 1$. The linear term in $\sqrt{-g}$ multiplies $R^{(1)}$ and consequently contributes only at $\mathcal{O}(h^3).$  Throughout this appendix all actions refer to the effective average action at scale $k$, we keep the subscript explicitly.

\subsection{Einstein--Hilbert term}

The quadratic part of the Einstein--Hilbert action reads
\begin{equation}
\Gamma^{(2)}_{k,\mathrm{EH}}
= \frac{Z_k}{2} \int d^4x \left(
- \frac{1}{2} h_{\mu\nu} \Box h^{\mu\nu}
+ \frac{1}{4} h \Box h
+ \ldots
\right),
\end{equation}
where the ellipsis denotes terms involving vector and transverse-–traceless tensor components, as well as gauge--dependent terms, which do not contribute to the scalar trace sector at quadratic in order in the quasi--static weak--field expansion and are therefore omitted. Projecting onto the trace yields the scalar
contribution
\begin{equation}
\Gamma^{(2)}_{k,\mathrm{EH,scalar}}
= \frac{Z_k}{8} \int d^4x h \Box h .
\end{equation}

\subsection{Curvature--squared terms}

We consider the curvature--squared part of the truncated action,
\begin{equation}
\Gamma^{(2)}_{k,R^2}
= \int d^4x \left(
a_k R^2 + b_k R_{\mu\nu} R^{\mu\nu}
\right).
\end{equation}
Expanding In the metric metric fluctuation $h_{\mu \nu}$,
\begin{equation}
(R^2)^{(2)} = 2 R^{(1)} R^{(1)}, \qquad
(R_{\mu\nu} R^{\mu\nu})^{(2)} = 2 R^{(1)}_{\mu\nu} R^{(1)\mu\nu}.
\end{equation}

Restricting to the scalar trace component and discarding total derivatives,
one finds
\begin{equation}
(R^2)^{(2)}_{\mathrm{scalar}} = (\Box h)^2, \qquad
(R_{\mu\nu} R^{\mu\nu})^{(2)}_{\mathrm{scalar}} = \frac{1}{3} (\Box h)^2 .
\end{equation}
Thus the scalar sector depends only on the single coupling combination
\begin{equation}
\alpha_k \equiv 3 a_k + b_k .
\end{equation}
The resulting scalar contribution from curvature--squared terms is
\begin{equation}
\Gamma^{(2)}_{k,R^2,\mathrm{scalar}}
= \frac{\alpha_k}{2} \int d^4x  h \Box^2 h .
\end{equation}

\subsection{Gauge fixing and ghost sector}

We employ de Donder--type gauge fixing. In the weak--field
expansion, neither the gauge--fixing nor the Faddeev--Popov ghost sector
contributes to the scalar trace kernel at quadratic order in the fields, evaluated up to the relevant derivative order. These sectors, therefore, do not affect the scalar projection considered here.

\subsection{Resulting scalar quadratic kernel}

Combining the Einstein--Hilbert and curvature--squared contributions, the
scalar part of the quadratic action takes the form
\begin{equation}
\Gamma^{(2)}_{k,\mathrm{scalar}}
= \int d^4x \left(
\frac{Z_k}{8} \, h \Box h
+ \frac{\alpha_k}{2} \, h \Box^2 h
\right)
+ \mathcal{O}(\partial^6).
\end{equation}
This expression defines the scalar quadratic kernel used in the main text.

\section*{Appendix B: Derivative Hierarchy and Scalar Self-Closure}
\label{app:derivative_hierarchy}
\setcounter{equation}{0}
\renewcommand{\theequation}{B.\arabic{equation}}

In this Appendix, we demonstrate that, under the quasi--static IR
scaling used in the main text, the scalar part of the quadratic kernel
closes at order $O(q^{4})$ in the Wetterich equation.  This requires
showing that: (i) vector and tensor components contribute only at
subleading orders $O(q^{4}\epsilon)$ and $O(q^{6})$ respectively, and
(ii) higher-derivative operators reduce to the scalar structures
already identified in Appendix~A.  Throughout, we use the York
decomposition on flat background (see e.g.\ Refs.~\cite{York1973})
and the quasi-static scaling defined in the main text.

\subsection*{B.1 Spatial derivative hierarchy}

The key object entering the FRG flow is the second functional
derivative $\Gamma^{(2)}_k$, which from Appendix~A has the scalar
structure
\begin{equation}
\Gamma^{(2)}_{k, scalar}
= Z_k \, \nabla^2
  + \alpha_k \, \nabla^4,
\label{eq:scalar-kernel-appendix}
\end{equation}

up to total derivatives and subleading terms.  To establish
self-closure, we examine possible scalar--vector and scalar--tensor
mixing terms after gauge fixing and projection onto the scalar trace.

\subsection*{B.2 Scalar--vector mixing}

In the standard York decomposition,
\begin{equation}
h_{\mu\nu}
=
h^{\rm TT}_{\mu\nu}
+
\partial_{(\mu} V_{\nu)}
+
\left(
  \partial_\mu \partial_\nu - \frac{1}{4}\eta_{\mu\nu}\Box
\right) \sigma
+
\frac{1}{4}\eta_{\mu\nu} h,
\end{equation}

Where $h^{\rm TT}_{\mu\nu}$ is the transverse–traceless tensor, $\sigma$ is the longitudinal scalar mode, and  the transverse vector is $V_\mu$. The divergenceless condition $\partial^\mu V_\mu = 0$ implies that any scalar--vector cross-term must involve at least one explicit derivative acting on the transverse vector $V_\mu$.  After projecting onto the scalar trace, the lowest possible non--vanishing scalar--vector mixing term is kinematic of the schematic form

\begin{equation}
h \partial_t \nabla \cdot V.
\end{equation}

Since $\nabla V \sim q V$ and $\partial_t \sim  \epsilon q$ and there are at least two extra spatial derivatives in the kinetic operator, derivative counting shows a leading order $O(\nabla^4 \epsilon)$ once the quasi--static scaling is imposed.
Thus, scalar--vector mixing enters only at

\begin{equation}
O(q^4 \epsilon) ,
\end{equation}

and therefore does not affect the scalar flow at $O(q^4)$.

\subsection*{B.3 Scalar--tensor mixing}

For the transverse-traceless (TT) sector, the conditions  $h^{\rm TT}=0$ and $\partial^\mu h^{\rm TT}_{\mu\nu}=0$ imply that a non--vanishing scalar contraction requires at least two additional derivatives. Any scalar--tensor mixing term must therefore contain four derivatives as the TT projection itself already carries two derivatives. Therefore, the lowest non--vanishing scalar--tensor contribution in the kinetic operator arises at the order $\nabla^6$. 
Consequently, the tensor sector contributes to the scalar flow only at $O(q^{6})$.  Therefore the tensor sector enters strictly beyond the order required for scalar closure.

We thus obtain the derivative hierarchy:
\begin{itemize}
 \item $\text{scalar} \sim O(q^2) + O(q^4),$
 \item $\text{scalar--vector} \sim O(q^4 \epsilon), $
 \item $\text{scalar--tensor} \sim O(q^6).$
\end{itemize}

\subsection*{B.4 Higher--derivative invariants}

The effective average action may contain additional operators such as
$R_{\mu\nu\rho\sigma}R^{\mu\nu\rho\sigma}$,
$R\,\Box R$, $R_{\mu\nu}\Box R^{\mu\nu}$, and analogous terms.  In four
dimensions, the Gauss--Bonnet combination ensures that the Riemann-squared invariant can be rewritten as a linear combination of $R^2$ and
$R_{\mu\nu}R^{\mu\nu}$ up to a topological surface term.  Operators with
six or more derivatives, such as $R\,\Box R$, contribute only to
$O(q^{6})$ or higher in the scalar sector. Consequently, within the quasi--static expansion truncated at $\mathcal{O}(q^4)$, such terms do not modify the scalar kernel and can be consistently neglected. 

\subsection*{B.5 Closure of the scalar sector}

With the above hierarchy, the Wetterich flow projected onto the scalar
trace is

\begin{equation}
\partial_k \Gamma_k^{\rm scalar}
= \frac{1}{2} \mathrm{Tr}
\left[
  \left(
    Z_k \nabla^2 + \alpha_k \nabla^4 + R_k
  \right)^{-1}
  \partial_k R_k
\right]
+ O(k^4 \epsilon, k^6),
\end{equation}

where $R_k$ is the regulator.  Since the vector and tensor sectors
enter only at $O(q^4 \epsilon)$ and $O(q^6)$ respectively, the flow of
the scalar couplings $Z_k$ and $\alpha_k$ is thus self--contained within
the scalar sub--sector at order $O(q^4)$.  The scalar sector, therefore
self-closes in the quasi-static weak-field IR limit, providing a consistent truncation of the full FRG equation.

This completes the derivation used in the main text.

\end{document}